\documentclass[aps,prb,twocolumn]{revtex4} %,preprint

\usepackage{graphicx}% Include figure files
\usepackage{dcolumn}% Align table columns on decimal point
\usepackage{bm}% bold math

\newcommand{\comment}[1]{}
\begin{document}
%\renewcommand{\theequation}{\arabic{section}.\arabic{equation}}
%\preprint{}   % Preprint number in upper right corner

\title{Is the Information Entropy
the Same as the Statistical Mechanical Entropy?}
\author{Phil Attard
\\ \texttt{phil.attard1@gmail.com} }
%\affiliation{\texttt{phil.attard1@gmail.com}}
%\\ \texttt{phil.attard1@gmail.com} }
%\email[]{E-mail: Phil.Attard1@gmail.com}
%\thanks{}
%
\date{17 September, 2012. arXiv:cond-mat} %Am.\ J.\ Phys.

\begin{abstract}
It is shown that the standard expression for the information entropy,
originally due to Shannon,
is only valid for a particular set of states.
For the general case of statistical mechanics,
one needs to include an additional term in the expression
for the entropy as a function of the probability.
A simple derivation of the general formula is given.
\end{abstract}

%\keywords{}

\maketitle

%%%%%%%%%%%%%%%%%%%%%%%%%%%%%%%%%%%%%%%%%%%%%%%%%%%%%%%%%%%%%%%%%%%%%%%%%%%%%%%
%                                                                             %
                \subsubsection{Introduction}
%                                                                             %
%%%%%%%%%%%%%%%%%%%%%%%%%%%%%%%%%%%%%%%%%%%%%%%%%%%%%%%%%%%%%%%%%%%%%%%%%%%%%%%

For a system that can exist in states $i=1,2, \ldots ,n$
with probability distribution
$ \underline p \equiv \{p_1, p_2, \ldots , p_n \}$
that is normalised, $ \sum_{i=1}^n p_i = 1$,
the information entropy is
\begin{equation} \label{Eq:plnp}
S[\underline p] = - k_\mathrm{B} \sum_{i=1}^n p_i \ln p_i .
\end{equation}
Boltzmann's constant, $k_\mathrm{B}$, is often suppressed
in informatic applications,
and the base of the logarithm is often taken to be 2.
The information entropy is a measure
of the disorder or lack of predictability of the system.

This formula for the entropy was originally introduced
by Shannon as a way of quantifying the information content
of messages.\cite{Shannon48,Shannon49}
(Actually the same formula was also given by Gibbs,\cite{Gibbs02}
who called it the average of the index of probability.
It also coincides with Boltzmann's H-function.)
The formula remains standard in information theory.
It is also widely used to calculate the entropy
in statistical mechanics.
Shannon's implicit assumption---%
that this information entropy
is the same as the entropy
that is used in thermodynamics and in statistical mechanics---%
has generally been accepted unquestioned,
and the formula has been used without modification in the physical sciences.
Jaynes based his maximum entropy formulation
of statistical mechanics
(wherein the appropriate probability distribution
is obtained by maximising the information entropy
subject to certain constraints)
on Eq.~(\ref{Eq:plnp}),\cite{Jaynes57}
and went on to use it as the basis of a novel interpretation
of logic and probability that has widespread ramifications.
\cite{Rosenkrantz83,Jaynes03}
It is also worth mentioning that statistical mechanical expansions
have been used to develop efficient methods for evaluating
the information entropy,  Eq.~(\ref{Eq:plnp}).
\cite{Attard97,Attard98,Attard99}

This note examines in detail Shannon's derivation
of Eq.~(\ref{Eq:plnp}),\cite{Shannon48,Shannon49}
which remains the standard text-book derivation.\cite{Jaynes03,Sethna11}
In Sec.~2
it is shown that the formula neglects the internal entropy of the states,
due to a certain confusion between the total entropy
and the entropy as a functional of the macrostate probability.
In Sec.~3 a simple derivation of the full result is reproduced,
\cite{Attard00,TDSM,NETDSM}
and in Sec.~4 it is discussed why this additional term
is essential to obtain the correct physical results
and agreement with the thermodynamic
and the  statistical mechanical entropy.

%%%%%%%%%%%%%%%%%%%%%%%%%%%%%%%%%%%%%%%%%%%%%%%%%%%%%%%%%%%%%%%%%%%%%%%%%%%%%%%
%                                                                             %
                \subsubsection{Derivation of the Information Entropy}
%                                                                             %
%%%%%%%%%%%%%%%%%%%%%%%%%%%%%%%%%%%%%%%%%%%%%%%%%%%%%%%%%%%%%%%%%%%%%%%%%%%%%%%

Following Shannon's original derivation
(see Appendix 2 of Ref.~\onlinecite{Shannon48},
or Refs~\onlinecite{Jaynes03,Sethna11}),
and changing only the notation
and adding a few clarifying remarks,
the derivation of the formula for the information entropy
is based on several self-evident axioms:
that it is a continuous function of the $p_i$,
that it increases with the number of states in the uniform case,
and that different ways of grouping the states must give the same value.

To understand this last axiom define a microstate
as a complete set of distinct, disjoint, and indivisible states,
and a macrostate as a set of microstates.
Let $i$ label the microstates used above, with probabilities $p_i$,
and let $I$ label the macrostates.
The macrostate $I$ contains $n_I = \sum_{i\in I}$ microstates,
and its probability is $q_I = \sum_{i\in I} p_i$.
The conditional probability of a microstate in a given macrostate is
$p_{i|I} = p_i/q_I , \; i \in I $,
which can be written as the $n_I$-component vector
$\{ p_i/q_I \} $.
In view of this definition,
the total entropy
can be written as the uncertainty due to the macrostate probability distribution,
plus the additional uncertainty due to locating the microstate
within each macrostate,
\begin{equation} \label{Eq:plnp-macro}
S[\underline p] =
S[\underline q] + \sum_I q_I  S[\{ p_i/q_I \} ] .
\end{equation}
The last axiom says that no matter how the microstates
are grouped into macrostates,
the total entropy must remain unchanged.

This equation is written in the standard form
(see, for example, Eq.~(11.8) of Ref.~\onlinecite{Jaynes03},
or the penultimate equation of Appendix 2 of Ref.~\onlinecite{Shannon48}),
but there is a subtle ambiguity with it that
ultimately causes the problem with the information entropy.
On the one hand, the notation implies that the entropy
expressed as a functional of the microstate probability
on the left hand side, $S[\underline p]$,
is the same function of its argument as the entropy
expressed as a functional of the macrostate probability
on the right hand side, $S[\underline q]$.
On the other hand, the left hand side of the equation
implies that the microstate form $S[\underline p]$ is the total entropy
of the system, whereas the right hand side implies that
the macrostate form $S[\underline q]$
is only part of the total entropy of the system.
The expression for the information entropy, Eq.~(\ref{Eq:plnp}),
results from overlooking this distinction,
as is discussed below.

Following  Shannon's Appendix 2,\cite{Shannon48}
or Jaynes' Eq.~(11.10),\cite{Jaynes03}
assume that the microstates are equally likely,
$p^\mathrm{u}_i = 1/n$,
so that the left hand side of Eq.~(\ref{Eq:plnp-macro}) is
\begin{equation}
S[\underline p^\mathrm{u}] = \sigma(n) .
\end{equation}
Further assume that the macrostates are all the same size,
$n_I = m$, so that $q_I = m/n$ and $ p_{i|I} = 1/m$.
Hence $S[\underline q^\mathrm{u} ] = \sigma(n/m)$
and $S[\{ p_i/q_I \} ] = \sigma(m)$.
With these two assumptions
Eq.~(\ref{Eq:plnp-macro}) becomes
\begin{eqnarray}
\sigma(n)  & = &
\sigma(n/m) + \sum_I q_I \sigma(m)
\nonumber \\ & = &
\sigma(n/m) + \sigma(m) .
\end{eqnarray}
This has evident solution
\begin{equation}
\sigma(n)  = k _\mathrm{B} \ln n .
\end{equation}
Apart from the choice of Boltzmann's constant and the logarithmic base,
this solution is unique.\cite{Shannon48,Jaynes03}
This result may be seen to be consistent
with Boltzmann's original definition of the entropy,
namely that the entropy of a state
is the logarithm of the number of molecular configurations in the state,
assuming that the configurations are all equally likely.

Now in order to deduce the form of the information entropy
as a functional of the probability distribution,
continue to take the microstates to be equally probable,
$p_i^\mathrm{u} = 1/n$,
but now take the the macrostate distribution to be non-uniform, $q_I \ne q_J$.
This is achieved by grouping different numbers of microstates $n_I$
into each macrostate $I$.\cite{Shannon48,Jaynes03}
In this case rearranging Eq.~(\ref{Eq:plnp-macro}) yields
\begin{eqnarray} \label{Eq:plnp-b}
S[\underline q]
& = &
S[\underline p^\mathrm{u}] - \sum_I q_I  S[\{ p_i^\mathrm{u}/q_I \}]
\nonumber \\ & = &
\sigma(n) - \sum_I q_I \sigma(n_I)
\nonumber \\ & = &
 - k_\mathrm{B} \sum_I q_I \ln  q_I  ,
\end{eqnarray}
since $q_I = n_I/n$.
With the apparently trivial replacement $I \Rightarrow i$,
this is the standard expression for the information entropy,
and so this appears to be a straightforward and unambiguous
derivation of Eq.~(\ref{Eq:plnp}).

However,
as noted above,
there is a  conceptual problem with Eq.~(\ref{Eq:plnp-macro}),
namely that the notation implies that
the microstate form $S[\underline p]$
and the macrostate form $S[\underline q]$
are the same functions of their arguments,
whereas the content of the equation
implies that $S[\underline p]$ is the total entropy
whereas $S[\underline q]$ is only part of the total entropy.
The only way to resolve this discrepancy
is to rewrite Eq.~(\ref{Eq:plnp-macro})
in terms of the macrostate probability
with the total entropy explicit,
\begin{equation} \label{Eq:plnp-macro-b}
S_\mathrm{total} =
S[\underline q] + \sum_I q_I  S[\{ p_i/q_I \} ] .
\end{equation}
The total entropy has to be independent
of how the microstates are grouped into macrostates, (Axiom 4).
Hence this result can also be used when the macrostates
are taken to be microstates themselves
(i.e.\ one microstate in each macrostate),
in which case this becomes
\begin{equation}
S_\mathrm{total} =
S[\underline p] + \sum_i p_i  S[\{ p_i/p_i \} ]
=
S[\underline p] + \sum_i p_i  \sigma(1)  .
\end{equation}
The simplest assumption is that there is no uncertainty
for a system consisting of a single microstate, $\sigma(1) = 0$.
(This assumption will be revisited in the next section.)
With these two equations, most of the preceding analysis holds.
Assuming a uniform distribution of both microstates
and macrostates one again finds that
$S_\mathrm{total} = \sigma(n) = k_\mathrm{B} \ln n$.
Assuming non-uniform macrostates, $q_I \ne q_J$,
and uniform microstates, $p_i^\mathrm{u} = 1/n$
does not change the total entropy,
$S_\mathrm{total} = k_\mathrm{B} \ln n$.
In this case Eq.~(\ref{Eq:plnp-macro-b}) may be rearranged as
\begin{equation} \label{Eq:plnp-q}
S[\underline q] =
k_\mathrm{B} \ln n - \sum_I q_I  \sigma(n_I)
= - k_\mathrm{B}  \sum_I q_I  \ln q_I .
\end{equation}
This has the appearance of the standard information entropy result,
but now Eq.~(\ref{Eq:plnp-macro-b}) shows explicitly
that this is only part of the entropy of the system.
Inserting this into Eq.~(\ref{Eq:plnp-macro-b}) gives the total entropy as
\begin{equation} \label{Eq:plnp-macro-c}
S_\mathrm{total} =
- k_\mathrm{B}  \sum_I q_I  \ln q_I  + \sum_I q_I  S_I  ,
\end{equation}
where in this case
the internal entropy of the macrostate
for equally likely microstates is
$S_I^\mathrm{u} = S[\{ p_i^\mathrm{u}/q_I \} ] = \sigma(n_I)$.
(This explicit result for the internal entropy of the macrostate
does not hold if the microstates are non-uniformly distributed,
but Eq.~(\ref{Eq:plnp-macro-c}) does hold in this general case,
as is shown in the next section.)

This formula remains predicated on the assumption
that the underlying microstates are uniformly distributed.
A derivation will be given in the following section without this restriction.
Before then however two comments can be made.
First, if the microstates are in reality  uniformly distributed,
$p_i^\mathrm{u}  = 1/n$,
then it turns out that the internal entropy of a microstate
can indeed be taken to be zero, $S_i^\mathrm{u} = 0$,
and this result for the total entropy expressed in terms
of microstates reduces to the information entropy form, Eq.~(\ref{Eq:plnp}).
In this case the result is trivial
\begin{equation}
S_\mathrm{total}^\mathrm{u} =
- k_\mathrm{B}  \sum_i p_i^\mathrm{u}   \ln p_i^\mathrm{u}
= k_\mathrm{B} \ln n .
\end{equation}
Because this result is trivial
(and in essence just a statement of Boltzmann's original definition
of the statistical mechanical entropy)
there really is no point in formulating the total entropy
in the information entropy form for uniformly distributed microstates.

Second, although both Eq.~(\ref{Eq:plnp-q}) and Eq.~(\ref{Eq:plnp-macro-c})
are true for non-uniformly distributed macrostates,
the derivation depends upon the microstates being uniformly distributed.
Additional assumptions or a different derivation is required when
the microstates are not uniformly distributed.
Similarly, there is nothing in the derivation
that says that either of these two equations can be applied
to non-uniformly distributed microstates
(i.e.\ $I \Rightarrow i$),
which Shannon\cite{Shannon48} and Jaynes\cite{Jaynes03} both assume
in their derivation of Eq.~(\ref{Eq:plnp}).

%%%%%%%%%%%%%%%%%%%%%%%%%%%%%%%%%%%%%%%%%%%%%%%%%%%%%%%%%%%%%%%%%%%%%%%%%%%%%%%
%                                                                             %
                \subsubsection{Derivation for Non-Uniform Microstates}
%                                                                             %
%%%%%%%%%%%%%%%%%%%%%%%%%%%%%%%%%%%%%%%%%%%%%%%%%%%%%%%%%%%%%%%%%%%%%%%%%%%%%%%

A simpler and more general
(in the sense that it applies for non-uniformly distributed microstates)
derivation of the entropy as a functional of the probability
distribution has been given by the author.\cite{Attard00,TDSM,NETDSM}
Suppose that the microstate $i$ has weight $w_i$.
These weights arise from the molecular details of the problem
and they need not be specified explicitly beyond the fact
that they are non-negative and linearly additive,
which are required for the system to satisfy the laws of probability.

Because the states are distinct, disjoint, and complete,
and because the weights are linearly additive,
the weight of a macrostate is $W_I = \sum_{i\in I} w_i$,
and the total weight is
\begin{equation}
W = \sum_i w_i = \sum_I W_I .
\end{equation}

The probability of a state is proportional to its weight,
\begin{equation}
p_i = \frac{w_i}{W}
, \mbox{ and }
q_I = \frac{W_I}{W} .
\end{equation}
These are obviously normalised.
The total weight is called the partition function in statistical mechanics.

Weight is the obvious  generalisation of number
in the case that the states are not equally likely.
Hence the entropy of a state
is defined to be Boltzmann's constant times the logarithm
of the weight of the state.
The microstate, macrostate, and total entropy are
\begin{equation}
S_i \equiv k_\mathrm{B} \ln w_i
, \;
S_I \equiv k_\mathrm{B} \ln W_I
, \mbox{ and }
S_\mathrm{total} \equiv k_\mathrm{B} \ln W,
\end{equation}
respectively.
It turns out that the entropy defined like this
has all the properties that one would desire of the entropy in physical systems.
The logarithm makes entropy an extensive variable,
like energy, number, and volume,
and this is what makes it so convenient for thermodynamics
and statistical mechanics.
%In particular, entropy is more convenient than weight
%because the total entropy of two independent states
%is just the sum of their individual entropies,
%whereas the total weight is the product of their individual weights.
%(Note the distinction between the composition of disjoint states,
%in which case weights are added,
%and the conjunction of joint states,
%in which case weights are multiplied,
%as in joint probabilities.
%The weight of a die showing an even number
%is the sum of the respective weights for 2, 4, and 6.
%The weight of a die showing 2 and a coin showing heads
%is the product of the respective weights.)

In view of the relationship between entropy and weight,
the probability of a state is proportional to the exponential
of the entropy divided by Boltzmann's constant,
\begin{equation} \label{Eq:pi=expSi/W}
p_i = \frac{e^{S_i/k_\mathrm{B} } }{W}
, \mbox{ and }
q_I = \frac{e^{S_I/k_\mathrm{B} } }{W} .
\end{equation}
This generic relationship between entropy and probability
is well known in statistical mechanics.
%An unappreciated consequence of this result for the information entropy
%will be discussed shortly.

Now to the major result,
namely the information entropy
for the case of non-uniform probability distributions.
In view of the normalisation of the probability distributions
it is straight forward to verify that
\begin{eqnarray} \label{Eq:S_tot(i)}
S_\mathrm{total}
& = &
k_\mathrm{B} \ln W
\nonumber \\ & = &
k_\mathrm{B} \sum_i p_i \ln W
\nonumber \\ & = &
k_\mathrm{B} \sum_i p_i \left[ \ln \frac{W}{ w_i } + \ln w_i \right]
\nonumber \\ & = &
- k_\mathrm{B} \sum_i p_i  \ln p_i + \sum_i p_i S_i .
\end{eqnarray}
In the above formulation there was no fundamental distinction
between macrostates and microstates,
and so the total entropy can equivalently be written as a functional
of the macrostate probability,
\begin{equation} \label{Eq:S_tot(I)}
S_\mathrm{total} =
- k_\mathrm{B} \sum_I q_I  \ln q_I + \sum_I q_I S_I .
\end{equation}
These are in the form of Eq.~(\ref{Eq:plnp-macro-c})
(for both microstates and macrostates)
but not in the form of the information entropy expression, Eq.~(\ref{Eq:plnp}).

%%%%%%%%%%%%%%%%%%%%%%%%%%%%%%%%
\subsubsection{Discussion}

This obvious contradiction between the information entropy result
Eq.~(\ref{Eq:plnp})
and the statistical mechanical entropy result,
Eq.~(\ref{Eq:S_tot(i)}), or  Eq.~(\ref{Eq:S_tot(I)}),
raises the question: when is it valid to neglect
the internal entropy of the states?

For the case that the microstates are equally likely,
one can indeed set their weight to unity, $w_i = 1$,
and their entropy to zero, $S_i = 0$.
But in this case Eq.~(\ref{Eq:plnp}) reduces to the trivial result,
$S_\mathrm{total} = k_\mathrm{B} \ln n$,
and there is no circumstance when the formula in terms
of the probability distribution is required.

In the case that the microstates are not equally likely,
one cannot neglect their internal entropy without explicit justification.
In statistical mechanics the generic cases where
the microstates are not equally likely  usually
involve hidden variables that are projected out of the problem.
One example is where one describes the microstates of the system
in terms of molecular coordinates
(e.g.\ the position and momenta of the center of mass of the molecules).
This neglect finer levels of description,
(e.g.\ the rotational coordinates, the bending and stretching of,
and the rotation about intramolecular bonds, the electron configuration, etc.).
Depending upon the specific system,
there may be good reason for neglecting
such finer levels of description,
and there may also be good reason for neglecting the internal entropy
due to them
(e.g.\ the internal configurations and their weight
might be the same for all the microstates,
and so the internal entropy might be constant).

The second example of the microstates having an internal entropy
is where the total system consists of a sub-system and a reservoir,
and the coordinates of the sub-system form the microstates.
For example,
the canonical equilibrium system
is that of a sub-system in contact with a thermal reservoir
of temperature $T$.
Here the microstates are the points in the phase space of the sub-system,
(the space of positions and momenta of the atoms of the sub-system).
However each sub-system phase space point ${\bf \Gamma}$
corresponds to multiple phase space points of the reservoir,
and the internal entropy associated with each phase space point
of the sub-system is $S({\bf \Gamma}) = -{\cal H}({\bf \Gamma})/T$,
where ${\cal H}$ is the Hamiltonian or total energy
of the sub-system.
This may be recognised as the change in entropy of the reservoir.
Neglecting this microstate entropy
by using the information theory expression, Eq.~(\ref{Eq:plnp}),
would give the wrong total entropy for the canonical equilibrium system.
With the Maxwell-Boltzmann distribution,
$\wp({\bf \Gamma}) = Z^{-1} \exp -{\cal H}({\bf \Gamma})/k_\mathrm{B} T$,
the information entropy form gives the total entropy as
\begin{eqnarray}
\tilde S_\mathrm{total}
& = &
- k_\mathrm{B} \int \mathrm{d}{\bf \Gamma} \,
\wp({\bf \Gamma}) \ln \wp({\bf \Gamma})
\nonumber \\ & = &
k_\mathrm{B} \ln Z
+ \frac{1}{T} \left< {\cal H}({\bf \Gamma}) \right> ,
\end{eqnarray}
whereas the full expression gives
\begin{eqnarray}
S_\mathrm{total}
& = &
\int \mathrm{d}{\bf \Gamma} \,
\wp({\bf \Gamma})
\left[- k_\mathrm{B}  \ln \wp({\bf \Gamma}) + S({\bf \Gamma}) \right]
\nonumber \\ & = &
k_\mathrm{B} \ln Z .
\end{eqnarray}
Clearly, only the full expression agrees with the well-known result
that the logarithm of the partition function
gives the total entropy of the system.
The information entropy expression gives the entropy of the sub-system alone,
neglecting the reservoir entropy.
Obviously, in seeking, for example, to optimise a system,
one should maximise the total entropy, not just part of it.

From this example, one can see explicitly
why the information entropy expression, Eq.~(\ref{Eq:plnp}),
is inappropriate for statistical mechanics.
The question remains: in what sense is it appropriate for information theory?

The information entropy singles out
a particular representation as the preferred representation,
namely the one in which $\sum_i p_i S_i = 0$.
Since only differences in entropy are significant,
it is always possible to add a constant to the entropy,
\begin{equation}
\tilde S_i = S_i + c ,\;
\tilde S_I = S_I + c , \mbox{ and }
\tilde S_\mathrm{total} = S_\mathrm{total} + c .
\end{equation}
Choosing $ c = - \sum_i p_i S_i $
(either explicitly or implicitly),
one sees that
\begin{eqnarray}
\tilde S_\mathrm{total} & = &
-k_\mathrm{B} \sum_i p_i \ln p_i
\nonumber \\ & = &
-k_\mathrm{B} \sum_I q_I \ln q_I
+ \sum_I q_I \tilde S_I .
\end{eqnarray}
The first equality is in the form of the information entropy,
but the second equality shows that
any shift to another set of states
requires the full expression for the total entropy.

In communications and informatic problems,
one generally does not have an underlying molecular description
of the message, signal, image, etc.,
and so one cannot give the actual value of the weight or the entropy
of a state.
However, one can measure the probability of a state
with relative ease.
The apparent advantage of Eq.~(\ref{Eq:plnp})
and of the first equality here
is that it depends only on the probability distribution,
not explicitly upon the entropy of the states.
Further, it is often the case that the microstates
are chosen based on the individual characters in a message
(or the pixels in an image),
and these are indivisible and arguably have no internal rearrangement
that can contribute to the information content of the message.
In this sense one can make a strong argument for applying  Eq.~(\ref{Eq:plnp})
in informatic applications \emph{provided}
that one explicitly restricts its use to such microstates.
One should not use that particular equation
in statistical mechanics
because it risks confusion between microstates and macrostates,
and also because it is not valid for the microstates
that typically appear in statistical mechanics.

%The expression for the total entropy,
%Eq.~(\ref{Eq:S_tot(i)}) is invariant
%to whether the system is described in terms of microstates
%or in terms of different collectives of macrostates.
%Invariance is an important scientific principle,
%and means, for example,
%that the equations of physics are
%insensitive to different systems of units
%or to different geometric representations.

%as simple as possible but no simpler

%\newpage %$\;$ \newpage \ \newpage
%{\bf \large References}
%%%%%%%%%%%%%%%%%%%%%%%%%%%%%%%%%%%%%%%%%%%%%%%%%%%%%%%%%%%%%%%%%%%%%%%%%%

%%%%%%%%%%%%%%%%%%%%%%%%%%%%%%%%%%%%%%%%%%%%%%%%%%%%%%%%%%%%%%%%%%%%%%%%%%
\end{document}